\documentclass[a4paper,10pt,twoside]{cpc-hepnp}

\usepackage{multicol}
\usepackage{graphicx}
\usepackage{amssymb,bm,mathrsfs,bbm,amscd}
\usepackage[tbtags]{amsmath}
\usepackage{lastpage}
\usepackage{epsfig}

\newcommand{\MMS}{M_{\rm rec}^2}
\newcommand{\yzero}{Y(4008)}
\newcommand{\y}{Y(4260)}

\newcommand{\yone}{Y(4360)}
\newcommand{\ytwo}{Y(4660)}

\newcommand{\BR}{{\cal B}}

\newcommand{\psp}{\psi(2S)}

\newcommand{\jpsi}{J/\psi}
\newcommand{\psift}{\psi(4040)}
\newcommand{\psifto}{\psi(4160)}
\newcommand{\psiftf}{\psi(4415)}
\newcommand{\EE}{e^+e^-}
\newcommand{\MM}{\mu^+\mu^-}

\newcommand{\kk}{K^+K^-}

\newcommand{\ppjpsi}{\pi^+\pi^- J/\psi}
\newcommand{\pppsp}{\pi^+\pi^- \psp}
\newcommand{\kkjpsi}{K^+K^- J/\psi}

\renewcommand{\arraystretch}{1.1}

\newcommand{\gisr}{\ensuremath{ \gamma_{ISR} }}

\newcommand{\RM}  {\ensuremath{ M_{\mathrm{rec}}   }}

\newcommand{\dd}  {\ensuremath{ D \overline D }}
\newcommand{\dpdm}{\ensuremath{ D^+ D^- }}
\newcommand{\ddb} {\ensuremath{ D^0 \overline D{}^0    }}

\newcommand{\ee}    {\ensuremath{ e^+e^- }}

\newcommand{\eeddg} {\ensuremath{ e^+e^- \to D \overline D \gamma_{ISR}}}
\newcommand{\eedpdm}{\ensuremath{ e^+e^- \to D^+ D^- }}

\newcommand{\eeddpg}    {\ensuremath{ e^+e^- \to D^0 D^- \pi^+ \gamma_{ISR} }}

\newcommand{\eeddnpig}{\ensuremath{ e^+e^- \to D \overline D (n)\pi \gisr  }}

\newcommand{\ddp}   {\ensuremath{ D^0 D^- \pi^+ }}

\newcommand{\eeddchg}  {\ensuremath{ e^+e^- \to D^{(*)+} D^{*-} \gisr}}

\newcommand{\eedpdstm} {\ensuremath{ e^+e^- \to D^+ D^{*-}      }}
\newcommand{\RMF} {\ensuremath{ M^\mathrm{fit}_{\mathrm{rec}} }}
\newcommand{\etal}{{\em et al.}}
\newcommand{\beq}{\begin{equation}}
\newcommand{\eeq}{\end{equation}}
\newcommand{\bitm}{\begin{itemize}}
\newcommand{\eitm}{\end{itemize}}

\begin{document}

\title{Study of Charmonium(-like) States via ISR at
Belle\thanks{Talk given at the BES-Belle-CLEO-BaBar Joint Workshop
on Charm Physics, Beijing, Chnia, November 26-27, 2007. Supported
by National Natural Science Foundation of China (10775142) and
100-talents program of CAS (U-25)} }

\author{%
YUAN Chang-Zheng$^{1;1)}$\email{yuancz@ihep.ac.cn} \\ (for the
Belle Collaboration) }

\maketitle

\address{
1~(Institute of High Energy Physics, Chinese Academy of Sciences,
Beijing 100049, China) }

\begin{abstract}

The cross sections for $\EE\to \ppjpsi$, $\pppsp$, $\kkjpsi$,
$D\overline{D}$, $\ddp+c.c.$, $D^*\overline{D}+c.c.$, and $D^*
\overline{D}{}^*$ are measured using data sample collected on or
near the $\Upsilon(4S)$ resonance with the Belle detector at KEKB.
A peak near 4.25~GeV/$c^2$, corresponding to the so called $\y$,
is observed in $\ppjpsi$ final state. In addition, there is
another cluster of events at around 4.05~GeV/$c^2$. Two resonant
structures are observed in the $\pppsp$ invariant mass
distribution, one at $4361\pm 9\pm 9$~MeV/$c^2$ with a width of
$74\pm 15\pm 10$~MeV/$c^2$, and another at $4664\pm 11\pm
5$~MeV/$c^2$ with a width of $48\pm 15\pm 3$~MeV/$c^2$. The rich
structures observed in all these final states indicate that our
understanding of the vector charmonium states above the open charm
threshold is still poor, let alone the other possible dynamics
such as charmonium hybrids or final state re-scattering and so on.

\end{abstract}



\footnotetext[0]{\hspace*{-2em}\small\centerline{\thepage\ --- \pageref{LastPage}}}%


\section{Introduction}

The study of charmonium states via initial state radiation ($ISR$)
at the $B$-factories has proven to be very fruitful. In the
process $\EE \to \gamma_{ISR} \ppjpsi$, the BaBar Collaboration
observed the $\y$~\cite{babary}. This structure was also observed
by the CLEO~\cite{cleoy} and Belle Collaborations~\cite{belley}
with the same technique; moreover, there is a broad structure near
4.05~GeV/$c^2$ in the Belle data. In a subsequent search for the
$\y$ in the $\EE \to \gamma_{ISR} \pppsp$ process, BaBar found a
structure at around 4.32~GeV/$c^2$~\cite{babar_pppsp}, while the
Belle Collaboration observed two resonant structures at
4.36~GeV/$c^2$ and 4.66~GeV/$c^2$~\cite{belle_pppsp}. Recently,
CLEO collected 13.2~pb$^{-1}$ of data at $\sqrt{s}=4.26$~GeV and
investigated 16 decay modes with charmonium or light
hadrons~\cite{cleoy4260}. The large $\EE\to \ppjpsi$ cross section
at this energy is confirmed. In addition, there is also evidence
for $\kk \jpsi$ (3.7$\sigma$) based on three events observed.
Belle also measured the process $\EE\to \kkjpsi$ via $ISR$ and
resonance-like structure was observed~\cite{bellekk}.

The total cross section for hadron production in $\EE$
annihilation in the energy region above the open-charm threshold
was measured by the Crystal Ball~\cite{cb:cs} and
BES~\cite{bes:cs} Collaborations. However, the parameters of the
vector charmonium states obtained from fits to the inclusive cross
section~\cite{seth,bes:fit} are poorly understood
theoretically~\cite{barnes}. Since interference between different
resonances depends on the specific final states, studies of
exclusive cross sections for charmed meson pairs in this energy
range are needed to clarify the situation. Recently, CLEO-c
performed a scan over $\sqrt{s}$ from 3.970 to 4.260~GeV and
measured exclusive cross sections for $D \overline{D}$, $D
\overline{D}{}^*$~\cite{foot} and $ D^* \overline{D}{}^*$ final
states at twelve points~\cite{cleo:cs}. A measurement of $\EE\to D
\overline{D}$ was performed at BaBar using the $ISR$
technique~\cite{babar:dd} with a much wider energy range. Belle
used a partial reconstruction technique to perform the
measurements of the exclusive cross sections including $\EE\to D
\overline{D}$~\cite{belle:ddb}, $D \overline{D}{}^*$, $ D^*
\overline{D}{}^*$~\cite{belle:dst}, and $D \overline{D}
\pi$~\cite{belle:psi} with the $ISR$ data.

The Belle analyses are based on a data sample of about
550~fb$^{-1}$ or 670~fb$^{-1}$ luminosity collected near the
$\Upsilon(4S)$ with the Belle detector~\cite{Belle} operating at
the KEKB asymmetric-energy $e^+e^-$ (3.5 on 8~GeV)
collider~\cite{KEKB}. About 90\% of the data were collected at the
$\Upsilon(4S)$ resonance ($\sqrt{s}=10.58$~GeV), and the rest were
taken at a center-of-mass energy that is 60~MeV below the
$\Upsilon(4S)$ peak.

\section{$\EE\to$ $h^+h^-$+charmonium}

Three final states are analyzed, including $\ppjpsi$, $\pppsp$,
and $\kkjpsi$. $\psp$ is reconstructed with its $\ppjpsi$ decays,
and the $\jpsi$ is reconstructed using its leptonic decays to
$\EE$ or $\MM$. All the charged tracks are required to be
positively identified as the particle species needed, and
$\gamma$-conversion events are further removed by particle
identification and invariant mass of the charged tracks. The
detection of the $ISR$ photon is not required, instead, we
identify $ISR$ events by the requirement on $\MMS$ close to zero,
where $\MMS$ is the square of the mass that is recoiling against
all the charged tracks.

\begin{figure}[htb]
\centerline{\psfig{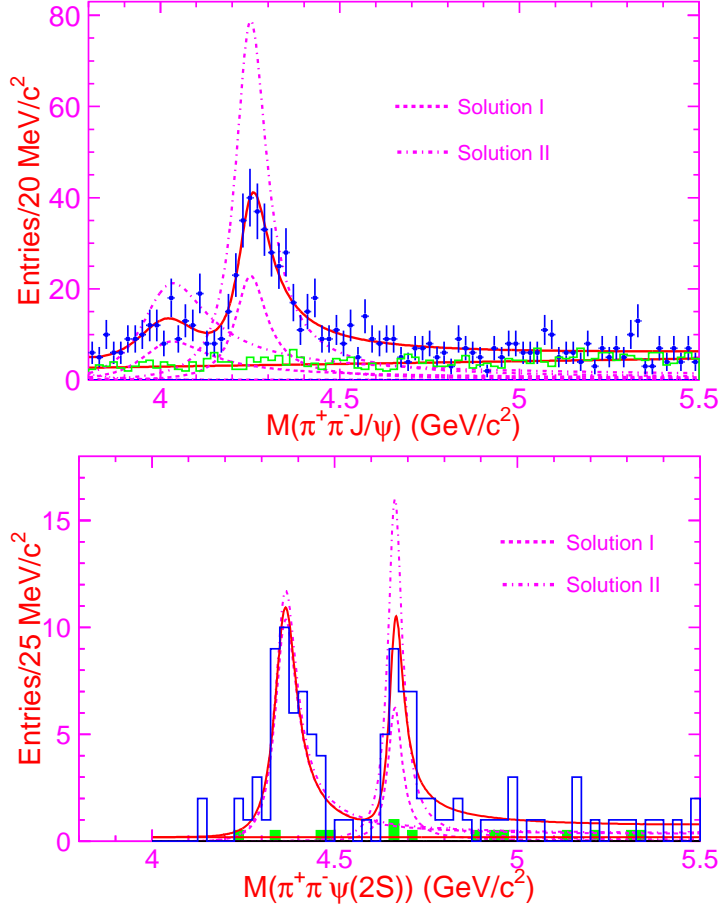}}
\caption{The $\ppjpsi$ (upper) and $\pppsp$ (lower) invariant mass
distributions and the best fit with two coherent resonances
together with a background term.} \label{mass}
\end{figure}

Figure~\ref{mass} shows the invariant mass distributions of
$\ppjpsi$ and $\pppsp$ after all the selection, together with a
fit with coherent resonance terms and a non-coherent background
term; and Fig.~\ref{xsection1} shows the resulting cross sections
for all the three final states, where the error bars indicate the
statistical errors only. Table~\ref{tab1} shows the fit results,
including the $\yzero$ and $\y$ from the $\ppjpsi$ mode, and the
$\yone$ and $\ytwo$ from the $\pppsp$ mode. It should be noted
that there are always two solutions in the fit to each mode, with
same mass and width for the resonances but with very different
coupling to $\EE$ pair ($\Gamma_{\EE}$). We also fit the $\kkjpsi$
invariant mass with resonances, but the statistics does not allow
us to discriminate the resonant structure.

\begin{figure}[htb]
\centerline{\psfig{file=hepnp1.epsi,height=12.0cm,angle=0,clip=}}
\caption{The measured $\EE\to \ppjpsi$ (a), $\pppsp$ (b), and
$\kkjpsi$ (c) cross sections.} \label{xsection1}
\end{figure}

\begin{table}[hbt]
\tabcaption{\label{tab1} Fit results of the $\ppjpsi$ and $\pppsp$
invariant mass spectra. The first errors are statistical and the
second systematic. $M$, $\Gamma_{\rm tot}$, and $\BR\cdot
\Gamma_{\EE}$ are the mass (in MeV/$c^2$), total width (in
MeV/$c^2$), product of the branching fraction to $\ppjpsi$ and the
$\EE$ partial width (in eV/$c^2$), respectively. $\phi$ is the
relative phase between the two resonances (in degrees).}
\renewcommand{\arraystretch}{1.2}
\begin{center}
\begin{tabular*}{100mm}{c@{\extracolsep{\fill}}ccc}
\hline\hline
  Parameters & ~~~Solution I~~~ & ~~~Solution II~~~ \\
  \hline
  $M(\yzero)$            & \multicolumn{2}{c}{$4008\pm 40^{+114}_{-28}$}  \\
  $\Gamma_{\rm tot}(\yzero)$   & \multicolumn{2}{c}{$ 226\pm 44\pm 87$}  \\
  $\BR\cdot \Gamma_{\EE}(\yzero)$
                  & $5.0\pm 1.4^{+6.1}_{-0.9}$ & $12.4\pm 2.4^{+14.8}_{-1.1}$  \\
  $M(\y)$            & \multicolumn{2}{c}{$4247\pm 12^{+17}_{-32}$} \\
  $\Gamma_{\rm tot}(\y)$   & \multicolumn{2}{c}{$ 108\pm 19\pm 10$} \\
  $\BR\cdot \Gamma_{\EE}(\y)$
                  & $6.0\pm 1.2^{+4.7}_{-0.5}$ & $20.6\pm 2.3^{+9.1}_{-1.7}$ \\
  $\phi$ [$\yzero$ and $\y$]          & $12\pm 29^{+7}_{-98}$ & $-111\pm 7^{+28}_{-31}$
  \\\hline
    $M(\yone)$            & \multicolumn{2}{c}{$4361\pm 9\pm 9$} \\
  $\Gamma_{\rm tot}(\yone)$   & \multicolumn{2}{c}{$74\pm 15\pm 10$} \\
  $\BR\cdot \Gamma_{\EE}(\yone)$
                  & $10.4\pm 1.7\pm 1.5$ & $11.8\pm 1.8\pm 1.4$ \\
  $M(\ytwo)$            & \multicolumn{2}{c}{$4664\pm 11\pm 5$} \\
  $\Gamma_{\rm tot}(\ytwo)$   & \multicolumn{2}{c}{$48\pm 15\pm 3$} \\
  $\BR\cdot \Gamma_{\EE}(\ytwo)$
                  & $3.0\pm 0.9\pm 0.3$ & $7.6\pm 1.8\pm 0.8$ \\
  $\phi$ [$\yone$ and $\ytwo$]          & $39\pm 30\pm 22$ & $-79\pm 17\pm 20$ \\
  \hline\hline
\end{tabular*}
\end{center}
\end{table}

\section{$\EE\to$ open charm}

Four final states are measured using $ISR$ data, they are $D
\overline{D}$, $D \overline{D} \pi$, $D \overline{D}{}^*$, and
$D^* \overline{D}{}^*$. $D^0$ candidates are reconstructed using
five decay modes: $K^- \pi^+$, $K^- K^+$, $K^- \pi^- \pi^+ \pi^+$,
$K^0_S \pi^+\pi^-$ and $K^- \pi^+ \pi^0$.  $D^+$ candidates are
reconstructed using the decay modes $K^0_S\pi^+$, $K^- \pi^+
\pi^+$ and $K^- K^+ \pi^+$. To improve the momentum resolution of
$D$ meson candidates, final tracks are fitted to a common vertex
applying the nominal $D^0$ or $D^+$ mass as a constraint.


The \eeddg\ signal events are selected by reconstructing both the
$D$ and $\overline D$ mesons, where $\dd=\ddb$ or \dpdm, and the
\gisr\ is not required to be detected; its presence in the event
is inferred from a peak around zero in the spectrum of the recoil
mass against the \dd\ system. To suppress background from
\eeddnpig\ processes we exclude events that contain additional
charged tracks that are not used in the $D$ or $\overline D$
reconstruction.


We select \eeddpg\ signal candidates in which the $D^0$, $D^-$ and
$\pi^+$ mesons are fully reconstructed and the \gisr\ is not
required to be detected as for $\dd$ mode, and a requirement on
the recoil mass squared against the \ddp\ system close to zero is
applied. Events contain additional charged tracks that are not
used in $D^0$, $D^-$ or $\pi^+$ reconstruction are removed to
suppress the background from $\ee \to D \overline D {(n)} \pi
\gisr ~({n}>1)$ processes.


The selection of \eeddchg\ signal events using full reconstruction
of both the $D^{*+}$ and $D^{*-}$ mesons, suffers from low
efficiency due to the low $D^{(*)}$ reconstruction efficiencies
and small branching fractions. Higher efficiency is achieved by
requiring full reconstruction of only one of the $D^{*}$ mesons,
the \gisr, and the slow $\pi_{\mathrm{slow}}$ from the other
$D^{*}$.

The analysis of the \eedpdstm\ is identical to that described
above for \eedpdm\ with the fully reconstructed $D^{*+}$ meson
replaced by a fully reconstructed $D^+$ meson. The requirement of
a detected slow pion from the unreconstructed $D^{*-}$ and a tight
requirement on $\Delta\RMF$ provides the clean \eedpdstm\ signal
peak in the distribution of $\RM (D^+\gisr)$.

\begin{figure}[htb]
\centerline{\psfig{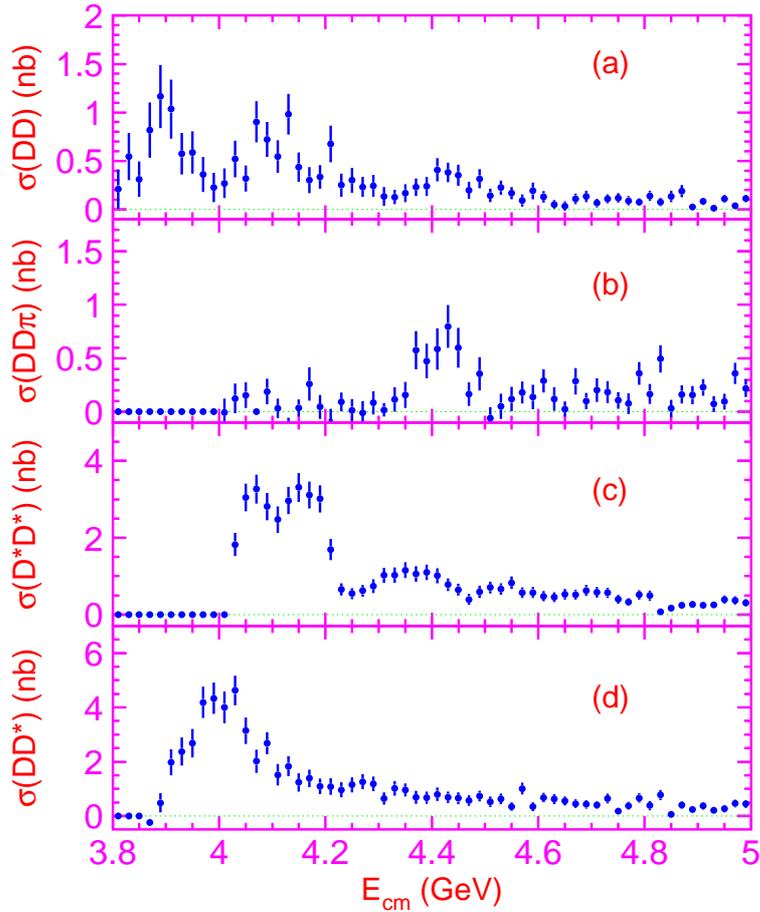}}
\caption{The measured $D \overline{D}$ (a), $D^0 D^- \pi^+ +c.c.$
(b), $D^{*+} D^{*-}$ (c), and $D^+ D^{*-}+c.c.$ (d) cross
sections.} \label{xsection2}
\end{figure}

The resulting cross sections of $\EE$ to the above modes are shown
in Fig.~\ref{xsection2}. We can see the structures at the $\psift$
and $\psiftf$ in $\dd$ mode, and a peak at 3.9~GeV which is in
qualitative agreement with the coupled-channel model prediction of
Ref.~\cite{eichten}; and $\psiftf\to \ddp$ is clearly seen in the
$\ddp$ final state. The shape of the $D^{*+} \overline{D}{}^{*-}$
cross section is complicated with several local maxima and minima,
especially large cross section near the $\psift$ and $\psifto$,
while in $D \overline{D}{}^*$ mode, aside from a prominent excess
near the $\psift$, the cross section is relatively featureless.

\section{Summary and concluding remarks}

With $ISR$ technique, Belle measured cross sections of $\EE\to
\ppjpsi$, $\pppsp$, $\kkjpsi$, $D \overline{D}$, $D \overline{D}
\pi$, $D \overline{D}{}^*$, and $D^* \overline{D}{}^*$ modes in
the center-of-mass energy between 3.8 and 5-6~GeV depends on final
states. These almost saturate the $\EE\to \rm{hadrons}$ production
cross section in this energy range. The structures observed in
these final states are very different from those observed in
inclusive hadrons~\cite{bes:cs,cb:cs}. This may suggest a very
different picture of the charmonium states in this energy range,
where the coupled-channel effect, final states rescattering,
threshold effect, and possibly existing charmonium-hybrids or
other exotic states such as tetra-quark state or molecular
state~\cite{steve,review} may contribute.

\section{Acknowledgments}

I congratulate the organizers for a successful workshop, and I
thank my colleagues at Belle for their wonderful work that are
presented in this talk; special thank to Galina Pakhlova for
supplying me the data for making Fig.~\ref{xsection2} in this
report.


\clearpage


\begin{thebibliography}{00}
\bibitem{foot} Charge-conjugate modes are included throughout this report.
\end{thebibliography}
\end{document}